# Tuning of large piezoelectric response in nanosheet-buffered lead zirconate titanate films on glass substrates

*Anuj Chopra[1], Muharrem Bayraktar[2,3], Maarten Nijland[1], Johan E. ten Elshof[1], Fred Bijkerk[3], and Guus Rijnders[1,]*\**

[1]Inorganic Materials Science Group, MESA+ Institute for Nanotechnology, University of Twente, PO Box 217, 7500 AE, The Netherlands

[2]Laser Physics and Nonlinear Optics Group, MESA+ Institute for Nanotechnology, University of Twente, PO Box 217, 7500 AE, The Netherlands,

[3]Industrial Focus Group XUV Optics, MESA+ Institute for Nanotechnology, University of Twente, PO Box 217, 7500 AE, Enschede, The Netherlands

Renewed interest has been witnessed in utilizing the piezoelectric response of PbZr$_{0.52}$Ti$_{0.48}$O$_3$ (PZT) films on glass substrates for applications such as data storage and adaptive optics. Accordingly, new methodologies are being explored to grow well-oriented PZT thin films to harvest a large piezoelectric response. However, thin film piezoelectric response is significantly reduced compared to intrinsic response due to substrate induced clamping, even when films are well-oriented. Here, a novel method is presented to grow preferentially (100)-oriented PZT films on glass substrates by utilizing crystalline nanosheets as seed layers. Furthermore, increasing the repetition frequency up to 20 Hz during pulsed laser deposition helps to tune the film microstructure to hierarchically ordered columns that leads to reduced clamping and enhanced piezoelectric response evidenced by transmission electron microscopy and analytical calculations. A large piezoelectric response of 280 pm/V is observed in optimally tuned structure which almost triples the highest reported piezoelectric response on glass. To confirm that the clamping compromises the piezoelectric response, denser films are deposited using a lower repetition frequency and a BiFeO$_3$ buffer layer resulting in significantly reduced piezoelectric responses. This paper demonstrates a novel method for PZT integration on glass substrates without compromising the large piezoelectric response.

## 1. Introduction

Perovskite oxides form a special and exciting class of materials which exhibit a wide spectrum of multifunctional physical properties such as superconductivity, photovoltaic effect, colossal magnetoresistance, ferroelectricity and piezoelectricity[1-5]. In particular, there has been an intensive research interest in using the piezoelectric response of ferroelectric thin films for a wide range of microelectromechanical systems (MEMS) such as in sensors and actuators[6-9]. Recently, an intriguing interest has been witnessed in integrating ferroelectric thin films on amorphous materials such as glass substrates for applications in data storage, electronic displays and adaptive optics[10-17]. These applications essentially demand high and stable piezoelectric response.

The piezoelectric response in thin films is known to depend on composition, growth quality, orientation and size of the fabricated devices[7,18-22]. Among all the ferroelectric materials $Pb(Zr_xTi_{1-x})O_3$ with morphotropic phase boundary composition, $Pb(Zr_{0.52}Ti_{0.48})O_3$ (PZT), is usually preferred, not only for its high piezoelectric response, but also for the large remnant polarization and low coercive electric field[5,9,23,24]. Therefore growth of well-oriented crystalline or epitaxial PZT films on glass substrates with control of the crystalline orientation is highly desired. Achieving epitaxial growth with (100) orientation is known to maximize the piezoelectric response compared to the polycrystalline films[18-19]. However the piezoelectric response is still drastically reduced compared to intrinsic piezoelectric response due to clamping of the thin film by the substrate[27-29]. On application of an electric field in the direction normal to the PZT film surface (longitudinal direction), the longitudinal expansion of the PZT film is coupled to a contraction in the direction parallel to its surface (transverse direction). Since the transverse contraction is constrained by the substrate, the effective longitudinal piezoelectric response of the PZT film is significantly reduced.

The clamping effect is less pronounced for structures that have small lateral dimensions, thus a lot of research has been driven to fabricate island-like nanostructures with submicron lateral size to enhance the piezoelectric response[21,30-32]. Indeed, experimentally it is probed that submicron size PZT capacitors have four times larger piezoelectric response in comparison to the large scale capacitors[31]. This larger piezoelectric response was later reported to be a primary manifestation of reduced clamping effect[32]. However, fabrication of island-like nanostructures and submicron size capacitors demand additional processing steps such as focused ion beam milling and chemical etching which increases the risk of contamination and damage. Thus, the reduced piezoelectric response of films that have lateral sizes relevant for MEMS devices (up to hundreds of microns lateral size), where the clamping effect is most dominant, still remains as a challenge.

Here we report on hierarchically ordered columnar growth of crystalline PZT films with preferred (100) orientation on glass substrates and minimization of the clamping effect by tuning the growth conditions. Locally epitaxial growth and control on growth orientation was achieved by utilizing crystalline nanosheets of $Ca_2Nb_3O_{10}$ (CNO) as seed layers on the glass substrates[33]. $LaNiO_3$ (LNO) electrodes and the PZT films were grown on the nanosheet-buffered glass substrates using pulsed laser deposition (PLD). PLD is a proven technique to grow high quality multifunctional oxide thin films and allows precise control on the growth quality by tuning of the deposition parameters such as laser fluence, repetition frequency of laser pulses, growth temperature and background gas pressure[34-37]. Most important for this work is the ability to change the microstructure of a film by changing the repetition frequency of the laser pulses. The following three heterostructures, named $H_{20Hz}$, $H_{5Hz}$ and $H_{BFO}$ hereafter, were deposited with different repetition frequency of the laser pulses. Within $H_{BFO}$, a $BiFeO_3$ (BFO) buffer layer is used.

$H_{20Hz}$: LNO (5 Hz) / **PZT (5 and 20 Hz)** / LNO (5 Hz) / CNO / Glass,

H$_{5Hz}$: LNO (5 Hz) / **PZT (5 Hz)** / LNO (5 Hz) / CNO / Glass,

H$_{BFO}$: LNO (5 Hz) / **PZT (5 Hz)** / BiFeO$_3$ (5 Hz) / LNO (5 Hz) / CNO / Glass.

## 2. Results and discussion

The XRD patterns reveal a predominantly (100)$_{pc}$ oriented PZT growth as shown in Figure 1(a) (shown only for H$_{20Hz}$). The subscript "pc" stands for the pseudo-cubic indexing which is used for all the materials in this paper. However a minor (110)$_{pc}$ reflection, which is two orders of magnitude smaller than the (100)$_{pc}$ peak, is also observed. The PZT films for all three heterostructures were found to have a pure perovskite phase without the presence of any impurity or pyrochlore phase. Only (100)$_{pc}$ orientation was observed for the LNO electrodes. It is also worth noting here that PZT films were also found in purely (100) orientation up to 750 nm[35]. The (100)$_{pc}$ oriented growth of the LNO bottom electrode is facilitated by the match of its lattice parameters to the underlying CNO nanosheets as schematically illustrated in Figure 1(b) and (c). The CNO nanosheets are known to have a 2D square lattice with a lattice parameter of $a_{CNO}$ = 3.86 Å which matches the in-plane lattice parameter of the LNO pseudo-cubic unit-cell as shown schematically in Figure 1(c)[38,39]. In this study, the lateral size of the CNO nanosheets deposited on glass substrates were around 2 μm[39]. Pure (100)$_{pc}$ oriented growth of the LNO bottom electrode facilitates a (100)$_{pc}$ oriented growth of the subsequent PZT layer. In addition to the XRD characterization, in-plane and out-of-plane crystal orientation of the PZT film for H$_{20Hz}$ was mapped using EBSD technique. The generated out-of-plane and in-plane pole figure maps are shown in Figure 2(a) and (b), respectively. The out-of-plane pole figure map in Figure 2(a) demonstrates a dominant (100)$_{pc}$ orientation (above 99% of the measured area) as almost the whole map has a single color (red color representing (100) orientation). A minor amount of (110)$_{pc}$ orientation (encircled region that has green color representing (110) orientation) is also observed, which is indeed in accordance to our XRD observations. The in-plane pole figure map in Figure 2(b) consists of predominately two colors representing the $(100)_{pc}$ and $(110)_{pc}$ orientations which signifies that the PZT film is randomly oriented in-plane. This in-plane random orientation of the PZT film is a manifestation of the random in-plane orientation of the CNO nanosheets lying underneath.

Detailed microstructure and thickness investigations were performed using transmission electron microscopy (TEM). A cross-sectional TEM image of a 2 µm thick PZT film on a 200 nm thick LNO bottom electrode for the $H_{20Hz}$ heterostructure is shown in Figure 3(a). The thicknesses of both top LNO and Pt electrodes were found to be ~100 nm. A magnified image was captured in order to analyze the interface quality between the glass substrate and the nanosheets as shown in Figure 3(b). A sharp and abrupt interface is visible between the glass and nanosheets, which confirms the effectiveness of using nanosheets as seed layers to grow a high quality crystalline LNO layer on an amorphous glass substrate. Further analysis of the microstructure for $H_{20Hz}$ heterostructure reveals that the PZT film can be distinctively identified into two regions along the growth direction as: (a) region 1 at the bottom where the PZT film has a dense structure, (b) region 2 at the top where the PZT film grows into separated columns. This difference in the microstructure of the PZT film was controlled by changing the frequency of the laser repetition. As discussed in the experimental section, first 250 nm of the PZT film was grown with a frequency of 5 Hz laser repetition which resulted in a densely packed growth as seen in region 1. After ensuring the full coverage of LNO bottom electrode with 5 Hz repetition frequency, the frequency was increased to 20 Hz which resulted in films with separated columnar structures as shown in region 2. On magnifying the separated columnar region (region 2) as shown in Figure 3(c), distinct columns with lateral sizes around 100 nm can be seen. The crystallography of these columns was analyzed using selected area electron diffraction (SAED). The SAED pattern for one of the columns that is shown in Figure 3(d) confirms an epitaxial growth.

For a clear comparison of the microstructures, cross-sectional TEM images of the three heterostructures are shown in Figure 4. The heterostructure $H_{20Hz}$ shown in Figure 4(a) has narrower (especially at the left half of the image) and more separated columns compared to other heterostructures. More importantly, the voids in $H_{20Hz}$ are penetrating much deeper into the PZT film compared to $H_{5Hz}$ shown in Figure 4(b). The heterostructure $H_{5Hz}$ still has a columnar microstructure in the upper half, but a non-columnar and continuous microstructure is visible in Figure 4(c) for $H_{BFO}$. The mechanism behind the transition from smaller and separated columnar growth in $H_{20Hz}$ to a wider and more connected columnar growth in $H_{5Hz}$ can be explained by considering the formation of islands during PLD. Guan et al. studied the effect of repetition frequency on the island size and island density using kinetic Monte-Carlo method and concluded that higher repetition frequency leads to formation of more and smaller islands as in our observation[40]. On the other hand, when the repetition frequency is lower, the islands have more time to become larger in lateral size. Similar results have been predicted and experimentally reported for $BaTiO_3$ films grown using PLD[41,42]. The transition from the partly columnar growth in $H_{5Hz}$ to continuous growth in $H_{BFO}$ can be explained by considering the lattice mismatch between the layers. The role of lattice-mismatch in controlling the quality of the films is also well studied[34,35,37]. The lattice parameters for the pseudo-cubic unit cells of LNO and PZT layers are $a_{LNO,pc}$ = 3.86 Å and $a_{PZT,pc}$ = 4.06 Å, respectively. The lattice mismatch between the PZT and LNO is 5.18 % [= $(a_{PZT,pc} - a_{LNO,pc}) / a_{LNO,pc} \times 100$] which results in a columnar growth as observed for PZT in heterostructures $H_{20Hz}$ and $H_{5Hz}$. In order to further increase the density of the structure, a 50 nm thick BFO layer was used as a buffer layer between the LNO and PZT films. The pseudo-cubic unit cell of BFO has a lattice parameter of $a_{BFO,pc}$ = 3.96 Å which reduces the lattice mismatch from 5.18 % (between PZT and LNO) to 2.52 % (between PZT and BFO) and hence promotes a denser PZT growth as evident from Figure 4(c). All in all, the TEM images in Figure 4 demonstrate that the columnar structure of the PZT film is dramatically influenced by the deposition parameters and can be tuned either by controlling the repetition frequency of the laser pulses or by using suitable buffer layers.

The measured longitudinal piezoelectric responses of the PZT films ($d_{33,f}$) are shown in Figure 5(a). A large effective piezoelectric coefficient of 280 pm/V was observed for the $H_{20Hz}$ heterostructure that has separated columns. However, much lower piezoelectric coefficients of 140 and 50 pm/V were measured for the $H_{5Hz}$ and $H_{BFO}$ heterostructures with denser microstructures, respectively. This large decrease in the effective piezoelectric response can be attributed to the increase in the clamping effect with increasing lateral size of the islands. The piezoelectric response in a thin film can be related to the intrinsic piezoelectric response ($d_{33}$) by[21,21,25]:

$$d_{33,f} = d_{33} - \frac{2s_{13,f}\sigma_f}{E_f}$$

(1)

where $s_{13,f}$ is the compliance of the film, $\sigma_f$ is the in-plane stress of the film that is causing the clamping and $E_3$ is the electric field applied in the out-of-plane direction. Considering that $d_{33}$, $s_{13,f}$ and $E_3$ are independent of the island size, $\sigma_f$ is the important parameter determining the dependence of $d_{33,f}$ to the island size. To explain better the in-plane stress, we consider the simplified geometry of the film and substrate shown in Figure 5(b). In this geometry, $w_f$, $h_f$ and $h_s$ are the half-width of the film island, thickness of the film and thickness of the substrate, respectively. An approximate formula of the in-plane stress as a function of $x$ (distance to the center of the island) has been derived by Suhir as[43]:

$$\sigma_f(x) = Y_f^0 \chi(x)\varepsilon_f .$$

(2)

The parameter $Y^0_f$ is the generalized Young's modulus[43] or the biaxial modulus[44] of the film that commonly replaces the Young's modulus ($Y_f$) to account for the two-dimensional stresses in thin film geometries. In our case, the film is anisotropic and has a columnar texture with columns aligned in the out-of-plane direction. For such films the generalized Young's modulus can be written as[45]:

$$Y^0_f = \frac{1}{s_{11,f} + s_{12,f}}. \tag{3}$$

The function $\chi(x)$ characterizes the distribution of the stress along the film width that is given by[43]:

$$\chi(x) = 1 - e^{-k(w_f - x)} \text{ with } k = \sqrt{\frac{3}{2}\left(\frac{s_{11,f} + s_{12,f}}{h_f}\right)\frac{Y_s}{h_s(1+v_s)}} \tag{4}$$

where $Y_s$ and $v_s$ are the Young's modulus and Poisson's ratio of the substrate, respectively. The parameter $\varepsilon_f$ is the in-plane strain of the film defined as $\varepsilon_f = d_{31}E_3$. To demonstrate the dependence of the effective piezoelectric response to the island width with a relevant example, we consider the theoretical elastic and intrinsic piezoelectric parameters of the tetragonal PbZr$_{0.5}$Ti$_{0.5}$O$_3$ material that has a composition very close to our films. The effective piezoelectric response ($d_{33,f}$) is calculated using the parameters[18,46,47] in Table 1 and plotted in Figure 5(b). For convenience, island half-width is replaced with island width in the figure. It is clearly visible from the figure that the effective piezoelectric response decreases from the intrinsic piezoelectric value to a lower, *i.e.* clamped, piezoelectric value as the island width increases. The effective piezoelectric response for an island width close to our column width (~100 nm) is 330 pm/V that is close to our measured piezoelectric response of 280 pm/V. The difference between the calculated and measured value is due to the clamped part of the columns at the bottom of the film (region 1 in Figure 3). The calculated piezoelectric response for the island width matching our continuous film width (200 μm) is 78 pm/V that is again close to our measured piezoelectric response of 50 pm/V. The reduced piezoelectric response of our films compared to calculated value can be due to additional clamping from the sides of the film which can cause effectively a larger island width and was not taken into account in the calculation. Lastly, the reason for the relatively smaller piezoelectric response of the $H_{5Hz}$ heterostructure (140 pm/V) can be explained by considering its microstructure. The separated columns in $H_{5Hz}$ heterostructure are visible only in the upper half of the film, which results in a significant reduction of the piezoelectric response due to the contribution from the lower continuous part. Here, we also remark that the effective piezoelectric coefficient of 280 pm/V is, to the best of our knowledge, the highest piezoelectric coefficient measured on glass substrates[33,49].

Last but not the least, for device applications of these films, a long-term switching stability on applying external electric field is essential. The stable operation of PZT films for all the three heterostructures was confirmed using fatigue measurements as shown in Figure 6. Up to the tested $5\times10^9$ operating cycles, the remnant polarization of the heterostructures $H_{20Hz}$ and $H_{5Hz}$ are stable, which demonstrate the applicability of these films in device applications for future technology. For the heterostructure $H_{BFO}$ a decrease in the remnant polarization is observed after $10^8$ cycles possibly due to BFO interface.

## 3. Conclusion

In summary, we demonstrated a novel approach to tune the piezoelectric response of PZT films by controlling the microstructure of the hierarchically ordered columns, without any chemical treatment. The microstructure was controlled either by changing the repetition frequency of PLD process or by introducing a suitable BFO buffer layer. PZT films deposited with 20 Hz repetition frequency showed separated columns which manifested a large piezoelectric response of 280 pm/V due to reduced substrate induced clamping effect. In case of films deposited with 5 Hz repetition frequency either directly on LNO or on BFO buffered samples, densely packed columnar or continuous growth resulted in a reduced piezoelectric response which is conclusively demonstrated the effect of the substrate clamping. To conclude, this work offers new possibilities to tune the piezoelectric response without any chemical treatment, which opens new avenues in thin film fabrication for the future device applications. The same approach can be extended to other oxide systems as well to tune their response.

## 4. Experimental methods

**Fabrication**

In order to promote crystalline growth and control the growth orientation of the subsequent layers, CNO nanosheets were deposited on glass substrates. The CNO nanosheets were exfoliated by chemical processing from their layer-structure parent compound $KCa_2Nb_3O_{10}$. In this process, the parent compound material $KCa_2Nb_3O_{10}$ was first treated with nitric acid to obtain a protonated compound which was then exfoliated to CNO nanosheets on further treatment with exfoliation agent tetrabutylammonium hydroxide. The exfoliated nanosheets were transferred to the glass substrates using Langmuir-Blodgett deposition process at room temperature. The glass substrates used in this article were of 500 μm thickness. Prior to the transfer of nanosheets, the glass substrates were first cleaned on a hot plate at 250 °C with a jet of supercritical $CO_2$ followed by oxygen plasma cleaning. More experimental details of the nanosheet preparation and deposition on glass substrates can be found elsewhere[38].

The nanosheet coated glass substrate was loaded into the PLD chamber and a base pressure of $5\times10^{-7}$ mbar was maintained before raising the substrate temperature. The nanosheet coated substrate was annealed at 600 °C for 60 minutes in 0.14 mbar oxygen pressure. All the oxide layers on the nanosheet coated glass substrates were deposited *in situ* by ablating materials from their respective stoichiometric targets using PLD with a KrF excimer laser operating at 248 nm wavelength with a pulse duration of 20 ns. Electrical measurements were facilitated using the LNO bottom and top electrodes. The LNO electrodes were deposited at 600 °C with a laser fluence and repetition frequency of 1.5 J/cm$^2$ and 5 Hz, respectively. For the PZT layer, PbZr$_{0.52}$Ti$_{0.48}$O$_3$ morphotropic phase boundary (MPB) composition was used to harvest the highest piezoelectric response. PZT films were deposited at 585 ºC in 0.27 mbar oxygen pressure with a laser fluence of 2 J/cm$^2$. During deposition of the PZT layer for the first heterostructure (H$_{20Hz}$), a dense PZT seed layer (~250 nm) was deposited with a frequency of 5 Hz repetition to avoid the short circuiting between the top and bottom electrodes. Then the frequency of laser repetition was switched to 20 Hz for the rest of the deposition. The PZT films were deposited completely with a frequency of 5 Hz laser repetition for both H$_{5Hz}$ and H$_{BFO}$ heterostructures. A BiFeO$_3$ (BFO) buffer layer was deposited with the same growth conditions as LNO layers for H$_{BFO}$ heterostructure. A 100 nm thick platinum (Pt) layer was deposited by radio frequency sputter deposition at room temperature on the top LNO electrode for all the heterostructures. The Pt layer deposited at the top improves the homogeneity of the electric field across the top electrode. Top electrodes with 200 × 200 μm$^2$ area were patterned using a standard photolithography process and structured by dry argon etching.

**Characterization**

The crystallographic properties of the samples were analyzed using an x-ray diffractometer (Philips X'Pert MRD) with Cu-*Kα* radiation. Pole figure maps were generated from electron back scattering diffraction (EBSD) patterns of the PZT films recorded using a high-resolution scanning electron microscope (HR-SEM, Zeiss MERLIN). The microstructure of the samples was analyzed by using transmission electron microscopy (TEM, Philips CM300ST - FEG operating at 300 kV). Mechanical and ion-beam etching based techniques were employed for sample preparation for TEM analysis. Fatigue measurements were recorded using a ferroelectric tester (TF analyzer 2000, aixACCT). The fatigue measurements were performed using fatigue pulses of 15 V amplitude with a frequency of 100 kHz. After each fatigue cycle, the *P-E* hysteresis loop was measured with a triangular pulse of 30 V amplitude at 1 kHz frequency. The macroscopic piezoelectric responses of the films were measured using a laser Doppler vibrometer (LDV, Polytec MSA-400) operating at 8 kHz. To carry out the LDV measurements, all the samples were glued to a large metal plate with silver paste to impede the bending of the substrates. The LDV measurements were performed by applying a small AC signal (0.5 V peak to peak amplitude) superimposed on a DC voltage sweeping from -12.5 V to +12.5 V.

**Acknowledgements**

This research program is funded by "Stichting Technologie en Wetenschap (STW)" under the contract 10448 with the project name "Smart Multilayer Interactive Optics for Lithography at Extreme UV wavelengths (SMILE)". The authors would like to thank Dr. Minh Nguyen for the Pt coating.

**Author Contributions**

A. C. and M. B. contributed equally to this article by performing the experiments and writing the text. M. N. prepared the nanosheet coatings. All the authors reviewed the manuscript.

**Additional Information**

**Competing financial interests:** The authors declare no competing financial interests.

**References**

[1] Baumert, B. A., Barium potassium bismuth oxide: A review. *J. Supercond.* **8**, 175-181 (1995).

[2] Alexe, M. & Hesse, D., Tip-enhanced photovoltaic effects in bismuth ferrite. *Nat. Commun.* **2**, 256 (2011).

[3] Jin, S. et al., Thousandfold Change in Resistivity in Magnetoresistive La-Ca-Mn-O Films. *Science,* **264**, 413-415 (1994).

[4] Vrejoiu, I. et al., Intrinsic ferroelectric properties of strained tetragonal $PbZr_{0.2}Ti_{0.8}O_3$ obtained on layer–by–layer grown, defect–free single–crystalline films. *Adv. Mater.* **18**, 1657-1661 (2006).

[5] Guo, R. et al., Origin of the High Piezoelectric Response in $PbZr_{1-x}Ti_xO_3$. *Phys. Rev. Lett.* **84**, 5423 (2000).

[6] Cross, L. E., Relaxor ferroelectrics. *Ferroelectrics,* **76**, 241-267 (1987).

[7] Trolier-McKinstry, S. & Muralt, P., Thin Film Piezoelectrics for MEMS. *J. Electroceram.* **12**, 7-17 (2004).

[8] Scott, J. F., Applications of Modern Ferroelectrics. *Science,* **315**, 954-959 (2007).

[9] Izyumskaya, N. et al., Processing, Structure, Properties, and Applications of PZT Thin Films. *Crit. Rev. Solid State Mater. Sci.* **32**, 111-202 (2007).

[10] Son, J. Y. & Shin, Y.-H., Highly c-Oriented $PbZr_{0.48}Ti_{0.52}O_3$ Thin Films on Glass Substrates. *Electrochem. Solid-State Lett.* **12**, G20-G22 (2009).

[11] Kim, D. H., Kim, Y. K., Hong, S., Kim, Y. & Baik, S., Nanoscale bit formation in highly (111)-oriented ferroelectric thin films deposited on glass substrates for high-density storage media. *Nanotechnology*, **22**, 245705 (2011).

[12] Roy, S. S. et al., Growth and characterisation of lead zirconate titanate (30/70) on indium tin oxide coated glass for oxide ferroelectric-liquid crystal display application. *Integr. Ferroelectr.* **29**, 189-213 (2000).

[13] Uprety, K. K., Ocola, L. E., & Auciello, O., Growth and characterization of transparent Pb(Zi,Ti)$O_3$ capacitor on glass substrate. *J. Appl. Phys.* **102**, 084107 (2007).

[14] Bruno, E. et al., Structural Transformations of PZT 53/47 Sol-Gel Films on Different Substrates Driven by Thermal Treatments. *Ferroelectrics*, **396**, 49-59 (2010).

[15] Bayraktar, M. et al., Active multilayer mirrors for reflectance tuning at extreme ultraviolet (EUV) wavelengths. *J. Phys. D: Appl. Phys.* **45**, 494001 (2012).

[16] Wilke, R. H. T. et al., Sputter deposition of PZT piezoelectric films on thin glass substrates for adjustable x-ray optics. *Appl. Opt.* **52**, 3412-3419 (2013).

[17] Bayraktar, M., Chopra, A., Rijnders, G., Boller, K. J. & Bijkerk, F. Wavefront correction in the extreme ultraviolet wavelength range using piezoelectric thin films. *Opt. Exp.* **22**, 30623-30632 (2014).

[18] Haun, M. J., Zhuang, Z. Q., Furman, E., Jang, S. J. & Cross, L. E., Thermodynamic theory of the lead zirconate-titanate solid solution system, part V: Theoretical calculations. *Ferroelectrics*, **99**, 63-86 (1989).

[19] Du, X., Zheng, J., Belegundu, U. & Uchino, K. Crystal orientation dependence of piezoelectric properties of lead zirconate titanate near the morphotropic phase boundary. *Appl. Phys. Lett.* **72**, 2421-2423 (1998).

[20] Chopra, A., Birajdar, B. I., Kim, Y., Alexe, M. & Hesse, D., Enhanced ferroelectric and dielectric properties of (111)-oriented highly cation-ordered PbSc$_{0.5}$Ta$_{0.5}$O$_3$ thin films. *J. Appl. Phys.* **114**, 224109 (2013).

[21] Nagarajan, V. et al., Dynamics of ferroelastic domains in ferroelectric thin films. *Nature Mater.* **2**, 43-47 (2003).

[22] Nagarajan, V., Scaling of the piezoelectric response in ferroelectric nanostructures: An effective clamping stress model. *Appl. Phys. Lett.* **87**, 242905 (2005).

[23] Scott, J. F. & de Araujo, C. A. P., Ferroelectric Memories. *Science*, **246**, 1400-1405 (1989).

[24] Schwarzkopf, J. & Fornari, R., Epitaxial growth of ferroelectric oxide films. *Prog. Cryst. Growth Charact. Mater*. **52**, 159-212 (2006).

[25] Lefki, K. & Dormans, G. J. M., Measurement of piezoelectric coefficients of ferroelectric thin films. *J. Appl. Phys*. **76**, 1764-1767 (19994).

[26] Xu, F., Chu, F. & Trolier-McKinstry, S., Longitudinal piezoelectric coefficient measurement for bulk ceramics and thin films using pneumatic pressure rig. *J. Appl. Phys*. **86**, 588-594 (1999).

[27] Chen, L., Li, J-H., Slutsker, J., Ouyang, J. & Roytburd, A.L., Contribution of substrate to converse piezoelectric response of constrained thin films. *J. Mater. Res*. **19**, 2853-2858 (2004).

[28] Ouyang, J., Ramesh, R. & Roytburd, A. L., Theoretical Predictions for the Intrinsic Converse Longitudinal Piezoelectric Constants of Lead Zirconate Titanate Epitaxial Films. *Adv. Eng. Mater.* **7**, 229-232 (2005).

[29] Prume, K., Muralt, P., Calame, F., Schmitz-Kempen, T. & Tiedke, S., Piezoelectric thin films: evaluation of electrical and electromechanical characteristics for MEMS devices. *IEEE Trans. Ultrason., Ferroelectr., Freq. Control,* **54**, 8-14 (2007).

[30] Roytburd, A. L., Alpay, S. P., Nagarajan, V., Ganpule, C. S. & Aggarwal, S., Measurement of Internal Stresses via the Polarization in Epitaxial Ferroelectric Films. *Phys. Rev. Lett.* **85**, 190-193 (2000).

[31] Bühlmann, S., Dwir, B., Baborowski, J. & Muralt, P., Size effect in mesoscopic epitaxial ferroelectric structures: Increase of piezoelectric response with decreasing feature size. *Appl. Phys. Lett.* **80**, 3195-3197 (2002).

[32] Nagarajan, V. et al., Realizing intrinsic piezoresponse in epitaxial submicron lead zirconate titanate capacitors on Si. *Appl. Phys. Lett.* **81**, 4215-4217 (2002).

[33] Bayraktar, M., Chopra, A., Bijkerk, F. & Rijnders, G., Nanosheet controlled epitaxial growth of $PbZr_{0.52}Ti_{0.48}O_3$ thin films on glass substrates. *Appl. Phys. Lett*. **105**, 132904 (2014).

[34] Leuchtner, R. E. & Grabowski, K. S., Ferroelectrics in *Pulsed Laser Deposition of Thin Films* (ed: Chrisey, D. B. & Hubler, G. K.) 473-508 (John Wiley & Sons Inc., 1994).

[35] Correra, L. & Nicoletti, S., Large-area deposition of thin films by UV pulsed laser ablation. *Mater. Sci. Eng. B*, **32**, 33-38 (1995).

[36] Zhu, Z., Zheng, X. J. & Li, W., Submonolayer growth of $BaTiO_3$ thin film via pulsed laser deposition: A kinetic Monte Carlo simulation. *J. Appl. Phys.* **106**, 054105 (2009).

[37] Chopra, A., Pantel, D., Kim, Y., Alexe, M., & Hesse, D., Microstructure and ferroelectric properties of epitaxial cation ordered $PbSc_{0.5}Ta_{0.5}O_3$ thin films grown on electroded and buffered Si(100). *J. Appl. Phys.* **114**, 084107 (2013).

[38] Nijland, M. et al., Local Control over Nucleation of Epitaxial Thin Films by Seed Layers of Inorganic Nanosheets. *ACS Appl. Mater. Interfaces,* **6**, 2777-2785 (2014).

[39] Chopra, A., Bayraktar, M., Bijkerk, F. & Rijnders, G., Controlled growth of $PbZr_{0.52}Ti_{0.48}O_3$ using nanosheet coated Si (001). *Thin Solid Films*, **589**, 13-16 (2015).

[40] Guan, L., Zhang, D. M., Li, X. &Li, Z. H., Role of pulse repetition rate in film growth of pulsed laser deposition. *Nucl. Instrum. Meth. B,* **266**, 57-62 (2008).

[41] Hinnemann, B., Hinrichsen, H. & Wolf, D. E., Unusual Scaling for Pulsed Laser Deposition. *Phys. Rev. Lett.*, **87**, 135701 (2001).

[42] Kim, D. H. & Kwok, H. S., Pulsed laser deposition of $BaTiO_3$ thin films and their optical properties. *Appl Phys Lett*, **67**, 1803−1805 (1995).

[43] Suhir, E., An Approximate Analysis of Stresses in Multilayered Elastic Thin Films. *J. Appl. Mech*, **55**, 143-148 (1988).

[44] Hsueh, C-H., Modeling of elastic deformation of multilayers due to residual stresses and external bending. *J. Appl. Phys*, **91**, 9652-9656 (2002).

[45] Huang, F. & Weaver, M. L., Effective biaxial modulus of ideally (hkl)-fiber-textured hexagonal, tetragonal, and orthorhombic films. *J. Appl. Phys*, **100**, 093523 (2006).

[46] Heifets, E. & Cohen, R. E., Ab initio Study of Elastic Properties of Pb(Ti,Zr)O$_3$. *AIP Conf. Proc*, **626**, 150-159 (2002).

[47] ULE® Corning Code 7972, Ultra Low Expansion Glass Datasheet (2006).

[48] Verardi, P., Dinescu, M., Craciun, F., Dinu, R. & Ciobanu, M. F., Growth of oriented Pb(Zr$_x$Ti$_{1-x}$)O$_3$ thin films on glass substrates by pulsed laser deposition. *Appl. Phys. A*, **69**, S837 (1999).

**Table 1.** Thickness, elastic and piezoelectric parameters of the film and the substrate used in the calculations.

| Material | $h$ [μm] | $Y$ [GPa] | $v$ | $s_{11}$ [1/TPa] | $s_{12}$ [1/TPa] | $s_{13}$ [1/TPa] | $d_{31}$ [pm/V] | $d_{33}$ [pm/V] |
|---|---|---|---|---|---|---|---|---|
| PbZr$_{0.5}$Ti$_{0.5}$O$_3$[18,46] | 2 | - | - | 5.76 | -0.28 | -4.98 | -156 | 330 |
| Glass substrate[47] | 500 | 67.6 | 0.17 | - | - | - | - | - |

**Figures**

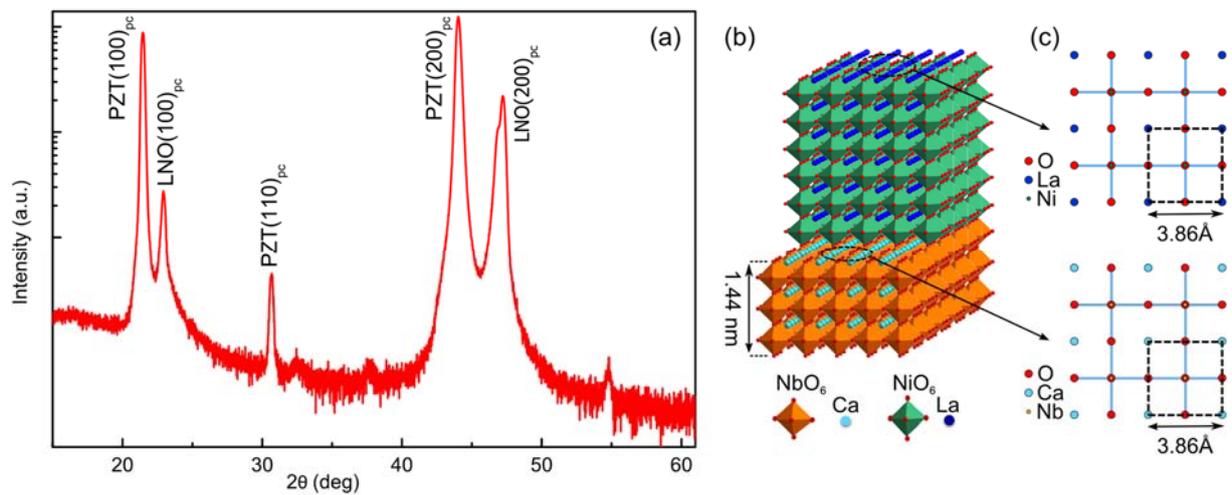

**Figure 1.** Crystal structure of the H$_{20Hz}$ heterostructure. (a) XRD *θ-2θ* scan. (b) Polyhedral representation of LaNiO$_3$ perovskite deposited on a perovskite-related Ca$_2$Nb$_3$O$_{10}$ nanosheet. (c) the square in-plane lattice of the LaNiO$_3$ and Ca$_2$Nb$_3$O$_{10}$. The ideal fitting of the lattice parameters resulting in (100)$_{pc}$ growth (subscript "pc" stands for pseudo-cubic indexing).

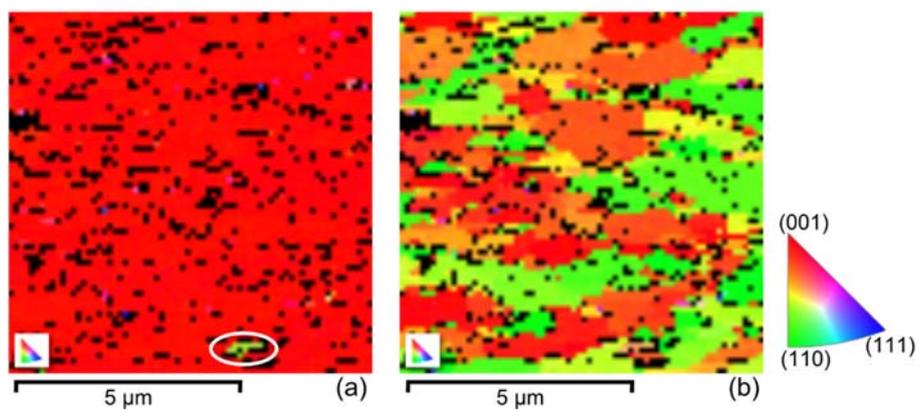

**Figure 2.** Pole figure maps in the (a) out-of-plane and (b) in-plane of the PZT film generated using electron backscatter diffraction measurements.

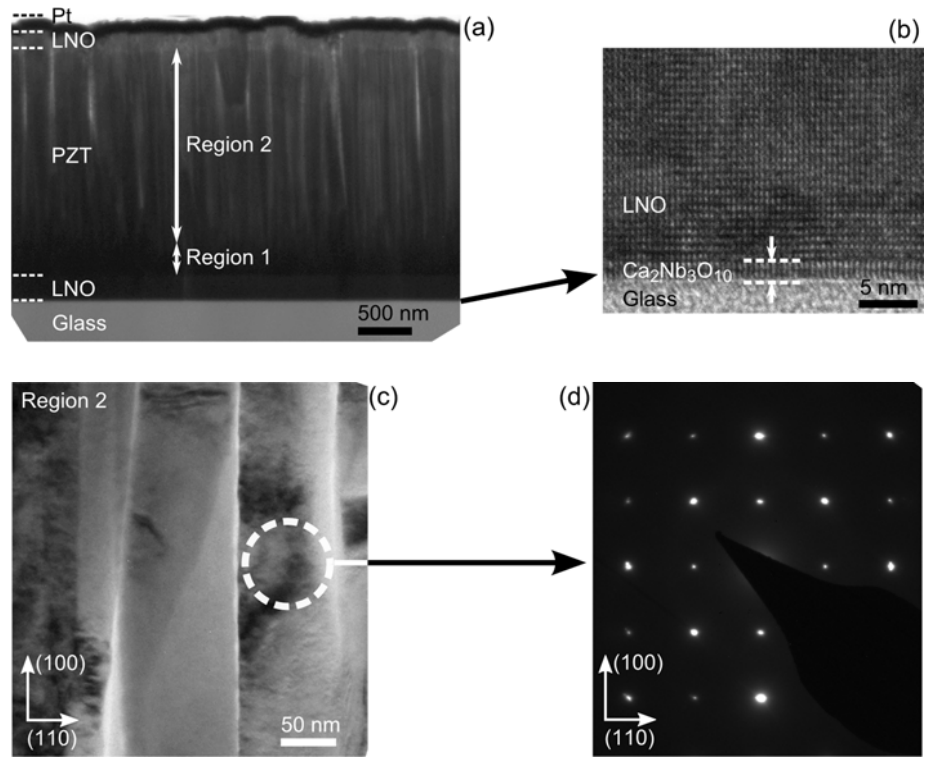

**Figure 3.** TEM images. (a) Cross-section of the H$_{20Hz}$ heterostructure in which PZT layer was first deposited at 5 Hz repetition frequency (region 1 with dense packing) and then 20 Hz repetition frequency (region 2 with separated columns). (b) A magnified image of Ca$_2$Nb$_3$O$_{10}$ nanosheet and glass interface. (c) A magnified image of the columns from region 2 of the PZT film. (d) Selected area electron diffraction pattern recorded for one of the column confirming an epitaxial growth.

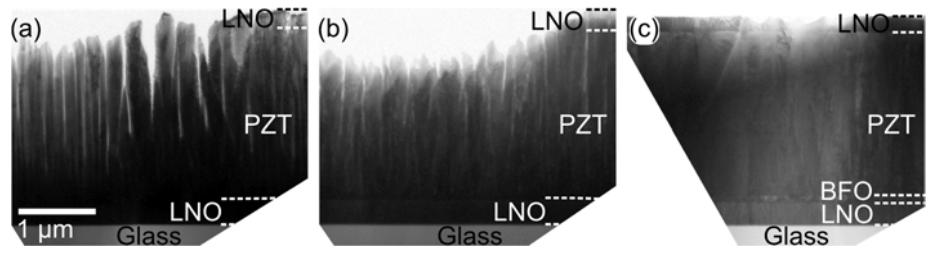

**Figure 4.** Cross-sectional TEM images revealing the impact of growth conditions on the heterostructures (a) H$_{20Hz}$, (b) H$_{5Hz}$, and (c) H$_{BFO}$.

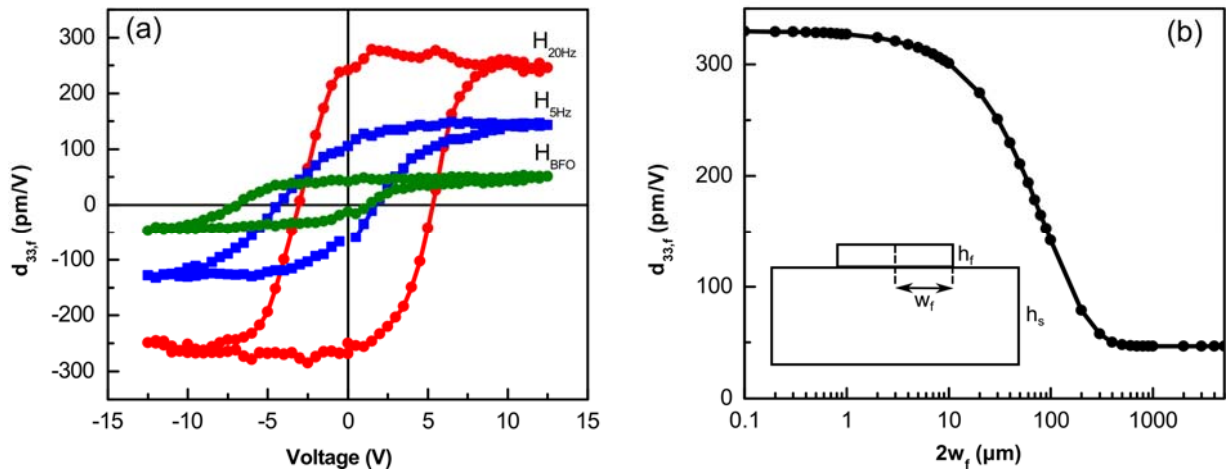

**Figure 5.** Piezoelectric response. (a) Longitudinal piezoelectric response ($d_{33,f}$) of the three heterostructures measured using a laser Doppler vibrometer. (b) Reduction of the piezoelectric response with increasing island width.

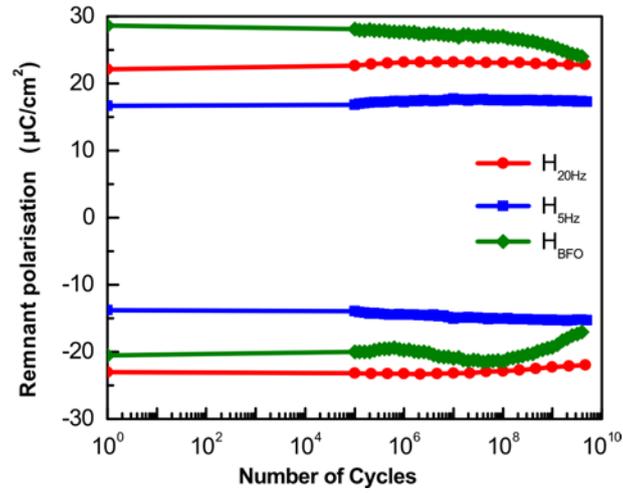

**Figure 6.** A plot of remnant polarization versus number of switching cycles.